# Evaluating the potential of thermoplastic polymers for cryogenic sealing applications: strain rate and temperature effects

Zhenzhou Wang[1*], Wendell Bailey[1], Junyao Song[1], Lingfeng Huang[1], Yifeng Yang[1]

1. Institute of Cryogenics, School of Mechanical Engineering, University of Southampton, Southampton SO17 1BJ, UK (*Corresponding author: Zhenzhou.Wang@soton.ac.uk)

**Abstract**

Cryogenic fuels, such as liquid hydrogen and liquid natural gas, emerge as versatile and sustainable energy carriers that are revolutionising various industries including aerospace, automotive, marine, and power generation. Thermoplastic polymers can be a suitable alternative to metal seals in cryogenic fuel systems. However, there is limited study about the behaviours of thermoplastics at cryogenic temperatures, especially at liquid hydrogen temperature of 20 Kelvin (K). This paper measured the tensile properties and coefficient of thermal expansion of three popular thermoplastics: PTFE, PEEK and UHMWPE at room temperature (RT), 77 K and 20 K and at four strain rates. Further microscopic analysis was also conducted to understand the failure mechanisms occurring when combining reduced temperature with varying strain rate. The tensile strength of each polymer increased from RT to 77 K and decreased from 77 K to 20 K. Young's modulus tended to increase, and the strain recorded at failure decreased when reducing temperature from RT to 20 K. From microscopic observation of PEEK and UHMWPE, a reduction in temperature from 77 K to 20 K resulted in a larger instantaneous fracture, with multi-faceted fracture surfaces containing many small "mirror like" and opaque or "misty" sub-regions within the fracture zone. For PTFE, the surface morphology exhibited an insensitivity to the increase in strain rate at cryogenic temperatures, and the microscopy showed how the size of dimples found within the fracture interface became smaller when temperature was reduced from 77 K to 20 K. Finally, PEEK was found to contract much less than PTFE and UHMWPE at 20 K, in agreement to it having the highest glass transition temperature of the three polymers, which is normally a good indicator when attempting to identify polymers that will tend to exhibit smaller contraction at cryogenic temperatures.

1. Introduction

Due to the increasing environmental legislation, industries encompassing aviation, maritime, automotive and power generation are actively transitioning their systems towards cleaner fuel sources to mitigate greenhouse gas emissions [1]. Alternative fuels, such as natural gas, which is primarily composed of methane and ethane, and hydrogen gas are being contemplated. Presently, the most efficient method of storing and transporting these alternative fuels is in their liquefied form, which are at cryogenic temperatures below 123 K [2]. The use of liquid hydrogen (20 K) coupled with liquid oxygen (90 K) as the oxidizer remains the optimal combination to fuel spacecraft [3].

To effectively utilise these cryogenic fuels and oxidizers on spacecrafts, aircrafts, ships and land-based vehicles; the construction of robust on-board fuel systems and fuel transportation pipelines is required. The selection of materials to create reliable seals at cryogenic temperatures is critical for the same applications. Cryogenic seals need to be gas tight even though they may primarily seal against liquid. The liquid will constantly boil to generate gas within the insulated tanks or transfer lines. Metal seals (copper, indium and aluminium) have been extensively used in cryogenic applications [2]. However, they require much greater forces to create a seal. Consequently, the size of flange becomes much thicker, and many bolts/clamps are required to apply necessary the clamping force. The classic knife edge seal design becomes less reliable when sealing against working pressures between 5-10 bar is required in combination with very low temperatures.

For applications that are highly sensitive to weight, such as those from aerospace and aviation, sealing materials lighter than metals could have advantages. Furthermore, metallic materials are susceptible to hydrogen embrittlement when they are exposed to a hydrogen-rich environment [1]. The combination of cryogenic temperature embrittlement and hydrogen embrittlement could further deteriorate the reliability of the seal [1,4]. Thermoplastic polymers present a potential alternative to metal seals in these applications, given their lower material cost, lighter weight, superior adaptability to complex shapes and reduced friction and wear in comparison to metals [5]. In addition, high-performance thermoplastic polymers, such as Polyether Ether Ketone (PEEK), are considered as possible liner choices for future cryogenic fuel tanks for launch vehicles due to their low hydrogen permeability, superior fracture toughness, damage tolerance and chemical resistance compared to thermosetting polymers [6,7]. Also composite materials containing a thermoplastic matrix are also considered as promising materials for future cryogenic propellant tanks [8,9].

However, overall there is a shortage of studies on the thermo-mechanical behaviours of thermoplastic polymers at cryogenic temperatures found in literature. The main focus of work currently found is centred on polymeric PEEK, and or composites that use PEEK as the matrix component [10–14]. From studies of thermoplastic polymer composites, most were proposed for aerospace applications [10–12,15]. Cui et al. [16] measured the tensile properties of four types of thermoplastics, but in thin film format at 77 K for the design of liquid oxygen hoses. More of the studies were carried out with the composites in form of the sheet. Chu et al. [10] presented a comparative study between the tensile behaviours of 30% chopped glass fibres reinforced PEEK composites at room temperature (RT), 77 K and 20 K. Their findings emphasised how the reinforced composites became stronger but also more brittle with decreasing temperature. They stated that this phenomenon was caused by a rapid increase in fibre-matrix interfacial shear strength when the temperature reduced. In another study on tensile properties, Aoki et al. [11] measured the tensile strengths of seven types of quasi-isotropic laminates at RT, 77 K and 4 K, including a study of IM600/PEEK carbon fibre reinforced thermoplastic. In contrast to Chu et al. [10], a reduction in tensile strength of up to 20% at cryogenic temperatures was reported compared to values measured at RT for all materials except the HTA/332 composite. They discussed how matrix cracks possibly led to the reduction in strength [11]. In addition, Adams and Gaitonde [12] examined the effects of temperature on the flexural modulus of unreinforced PEEK, HTA, and carbon fibre reinforced composites. Similar to Chu et al, their results found the stiffness decreased as temperature increased for all materials, with PEEK displaying a near-linear decrease [12]. Unexpectedly, the HTA matrix composite had a dramatic drop in stiffness at about 200 K [12]. Gates et al. [15] investigated the mechanical properties of a carbon fibre polymeric composite, IM7/PETI-5, with respects to temperature and aging. Their study reported the longitudinal and transverse stiffnesses and strengths under tension and how they decreased as the temperature decreased [15].

Although tested under the same conditions, disagreements are found comparing the main results and trends of Chu et al. [10], and Adams and Gaitonde [12] with those of Aoki et al. [11] and Gates et al. [15]. The former set of studies concluded an increase in strength for composites at low temperatures, whereas the latter studies reported a decrease in stiffness and strength for their composites. This discrepancy is found repeatable across literature has also been expressed by Wang et al. in their review paper [2].

From the review of existing work investigating the thermal properties of polymer at cryogenic temperatures, Fabian et al. [16] measured the coefficient of thermal expansion (CTE), and thermal conductivity of four types of thermosetting polymers, and one type of thermoplastic polymer (Polyimide), from 295 K to 4 K. Their research was undertaken to aid the design of insulation systems for the toroidal field coils for the International Thermonuclear Experimental Reactor (ITER) project. They found Polyimide contracted less than bismaleimide and TGDDM epoxies [16]. Chu et al. [10] measured the thermal properties of thermoplastic polymers and thermoplastic matrix composites down to 77 K and found the CTE of PEEK matrix composite and unfilled PEEK were not influenced by reduced temperatures from RT to 77 K. In contrast, Adams and Gaitonde [12] also measured the CTE of AS4/HTA laminate (another thermoplastic matrix composite) between RT and 120 K and reported an increase in the CTE with respects to reducing temperature.

After surveying the current state-of-art, it is clear the limited number of studies and contradictory conclusions and trends evaluated from existing studies are hindering the wider use of thermoplastics at cryogenic temperatures, especially in retrospect to the growing demands to find reliable solutions for storing liquid hydrogen at 20 K. Gates et al. [15] mentioned that material tests should be executed at the temperature of interest since the changes in stiffness and strength with respects to temperature are non-linear. There is necessity for rigorous testing to resolve conflicting results and enrich the database of cryogenic test of polymeric materials. Therefore, this paper aims to investigate the tensile properties and CTE of three types of popular thermoplastic polymers at RT, 77 K and 20 K. These include polytetrafluoroethylene (PTFE), polyether ether ketone (PEEK) and ultra-high molecular weight polyethylene (UHMWPE). For tensile testing, although it is commonplace to use RT test standards as reference to guide the cryogenic test approach; question marks remain over the use of the fast test speeds when conducting cryogenic tests [2]. Furthermore, high strain rate significantly affects the mechanical properties of polymers because of their viscoelastic nature [5]. Therefore, this study will also investigate the strain rate dependency upon the tensile properties of the same polymers at the same three test temperatures. It is anticipated for this work to benefit the wider implementation of these polymers, and help guide strain rate selection for future cryogenic testing of other polymeric materials.

This paper is presented in five sections. Section 2 provides details of the experimental setups. Section 3 summarises the results of tensile tests and thermal expansion measurements and with microscopy observation describes the fracture surfaces under different test conditions. Section 4 discusses the effects of temperature and strain rate on the material behaviours at both macro- and micro- levels. The entire study is concluded in Section 5.

## 2. Experimental procedures

A test programme was conducted to investigate the strain rate dependency on the tensile stress-strain mechanical data, and the thermally induced CTE of three types of thermoplastic polymers under cryogenic temperatures. The details of the test coupons, cryogenic tensile test set up, cryogenic CTE measurement set up are described hereafter.

2.1 Tensile test set-up

Three batches of thermoplastic polymer tensile coupons were prepared (PTFE, PEEK and UHMWPE). Since no specific test standard exists to guide cryogenic tensile testing of thermoplastic polymers, ISO 37 [17] was selected as the guideline for the test method and setup, since it specifically refers to testing of thermoplastics at room temperature compared to ISO 527-1 [18]. Tensile coupons were designed in accordance to the Type I dumbbell profile described in ISO 37 [17], and cut from the flat sheets using waterjet to the nominal dimensions shown in Figure 1. The three polymers were tested under tension at RT, 77 K and 20 K. According to ISO 37 [17], the recommended test speeds range from 0.125 mm/min to 500 mm/min. Since most cryogenic test studies of polymers and polymeric composites are normally executed at 2 mm/min [2], this study used this speed as a baseline and tested the coupons at a five times increase of test speeds. The tests were conducted at 2 mm/min (0.00159/s), 10 mm/min (0.00793/s), 50 mm/min (0.03968/s) and 250 mm/min (0.19841/s) of crosshead displacement speeds.

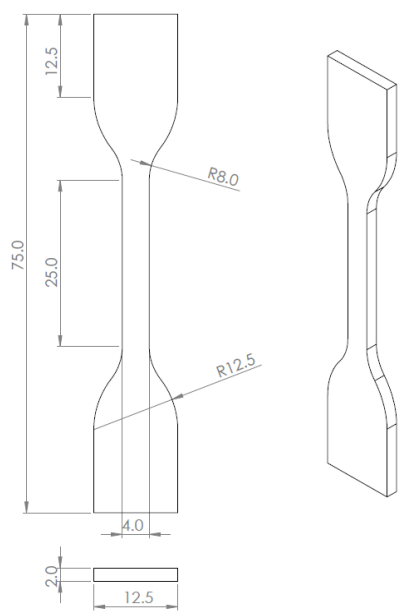

Figure 1. The nominal dimensions of test coupons.

The tensile tests of thermoplastic were conducted on a servo-mechanical Instron 3369 test machine. The load applied on the test coupons were performed with a calibrated load cell working in a suitable load range to cover both the RT and cryogenic tests. A bespoke testing platform was designed to fit onto the machine shown in Figure 2a. The testing platform was fixed to the base of the machine. The loading bar was attached to the load cell and could pass freely through the hole in the top plate of the support frame. Two pillars supporting a "floating stage" were fixed to the same top plate. Each end of the coupon was clamped by a pair of grips, shown in Figure 2b. The top grip was connected to the bottom of the loading bar and the bottom grip was fixed to the floating stage.

The cutaway profile (negative) leftover after waterjet cutting out the dumbbell was replicated and cut out from a sheet of stainless steel with the same thickness as the coupon. This "negative" profile was used to locate and restrain the coupon during the tensile test as illustrated in Figure 2b. The pair of negatives especially worked well to hold the test coupon in position at cryogenic temperatures. They prevented the coupon from slipping between the flat faces of the grips as the clearance between them increased during cooling down. The strain measurements were performed with a calibrated 10 mm gauge length Epsilon Model 3442 clip-on extensometer. The lowest working temperature range of this particular model of extensometer is 4 K.

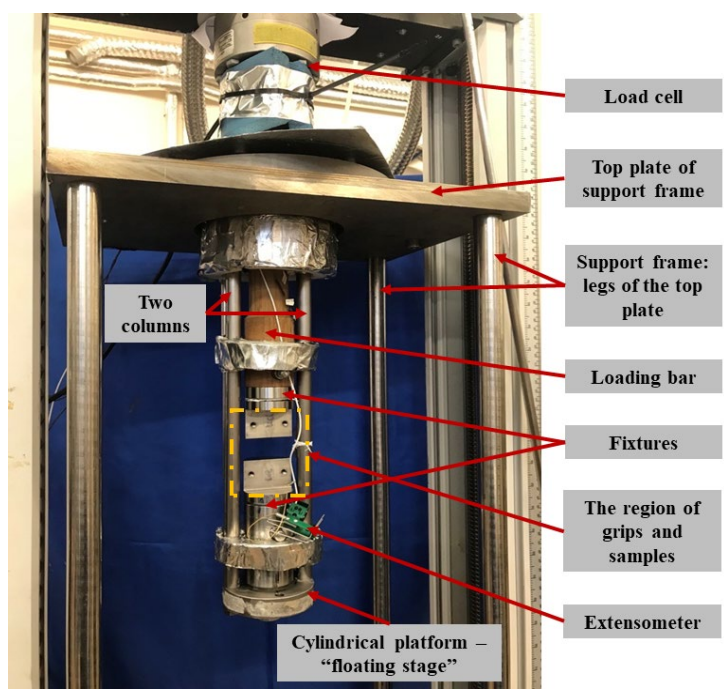

(a)

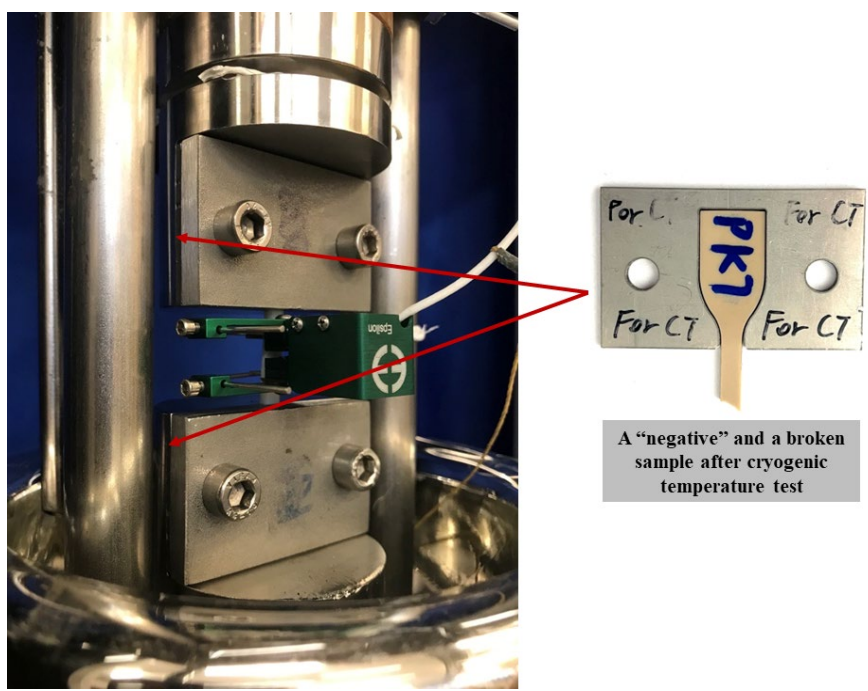

(b)

Figure 2. The figure of bespoke designed testing platform: (a) support frame; (b) details of sample region.

For the cryogenic tensile tests, a double-walled vacuum insulated silver-coated glass Dewar was inserted from underneath like a sleeve to surround the floating stage and grips to form a cold gas/liquid containing envelope. The Dewar was filled with either liquid nitrogen to form a bath to submerge the coupon at 77 K, or filled with cold helium gas at 20 K and relied on gas pressure and baffles to tubulise the flow to efficiently cool the coupon by convection as shown in Figure 3a. For test at 77 K, the grip, floating stage and pillars were pre-cooled directly in liquid nitrogen before mounting the coupon and extensometer. This process reduced the pre-load generated by the shrinkage of the entire rig imposing significantly on the grips and the coupon. After reaching 77 K, the sample was quasi-statically held at this temperature for 10 minutes before executing the tensile test.

For 20 K tests, cryogenic helium gas was boiled using a heater placed directly inside the liquid helium Dewar. A sketch of the entire helium gas cooling system is illustrated in Figure 3c. A double-walled vacuum insulated transfer line was inserted into the Dewar to collect the cold helium gas near the liquid surface, which was close to 6 K. The gas was delivered into the bottom of glass Dewar to cool the walls of the glass vessel, floating stage, grips and coupon. During the initial stage of the cooling, a gas helium bottle was connected to the liquid helium storage vessel to increase the internal pressure of the storage vessel to maintain a steady gas flow. Because cooling requires good heat exchange between the helium gas and test coupon; three baffles were placed inside the glass Dewar to optimise convection by creating a tortuous path for the gas flow. The temperature of test coupon was measured and recorded by three calibrated Temati (carbon ceramic) thermometers shown in Figure 3b. Two of the three thermometers were fixed directly on the top and bottom grips. The other thermometer was placed in the vicinity of the coupons gauge region. When this thermometer reached 20 K, the temperature gradients between this thermometer and those placed at the grips were within ±2 K. A temperature gradient of ±5 K is very reasonable to obtain repeatable results at cryogenic temperature. For each coupon, the cooling process required approximately 4 to 5 hours to reduce the temperature from RT to 20 K.

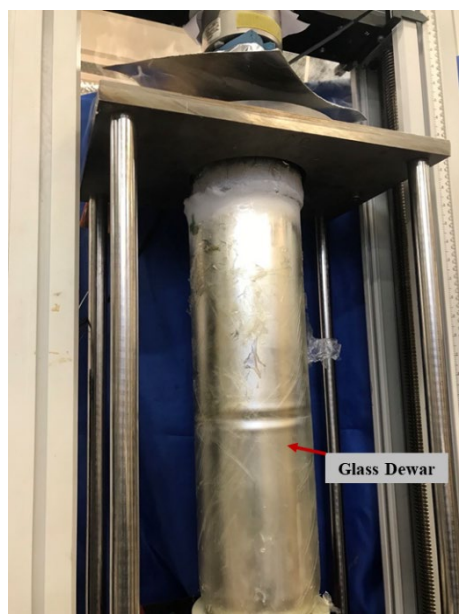

(a)

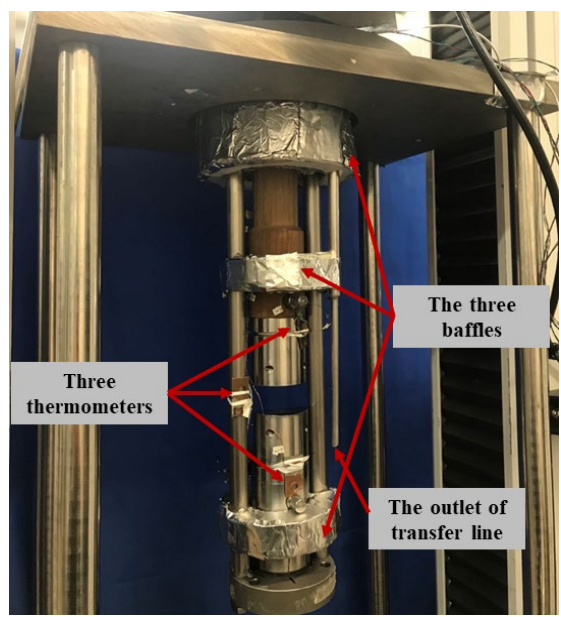

(b)

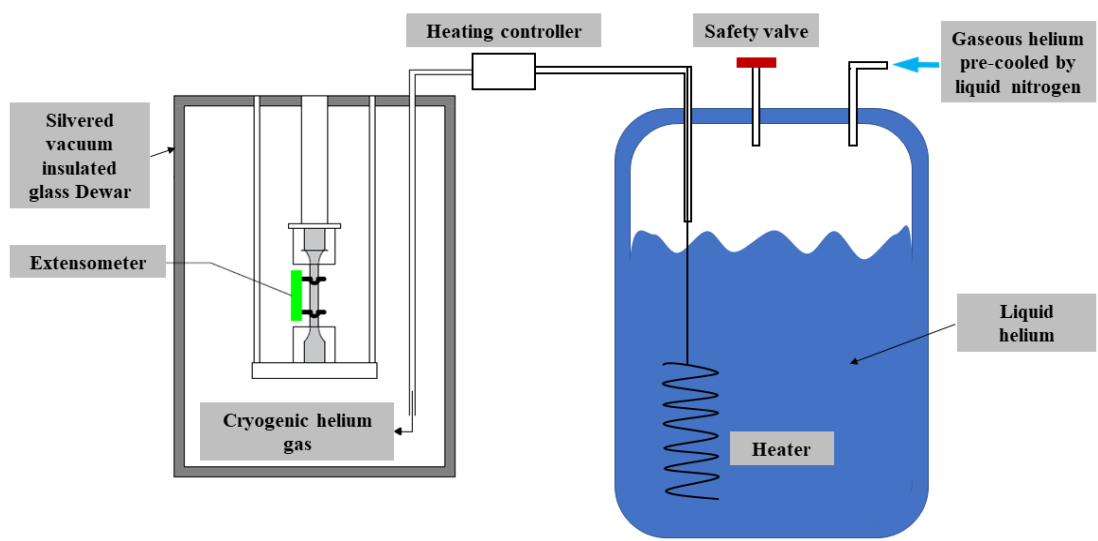

(c)

Figure 3. The setup of cryogenic chamber: (a) lifted glass Dewar; (b) thermometers, baffles and the transfer line; (c) a sketch of the helium gas cooling system.

The Instron's BlueHill 3 software simultaneously recorded the load, displacement of crosshead, and the strain measured by the extensometer during the test. A separate data logger recorded the temperature of the coupon. The tensile modulus was calculated using the slope of stress-strain curve between 0.0005 and 0.0025 strains in accordance to the ISO 37 [17]. The ultimate tensile stress (UTS), strain at UTS, and strain at break were also calculated in accordance to the same standard. Since the measurements of PTFE, PEEK and UHMWPE exceeded the strain range of the extensometer during RT tests, the nominal strain was calculated using the sample grips displacement (recorded from the movement of the machines crosshead divided by the starting distance between the grips (21 mm)).

The tensile modulus of the coupons was still obtained from the stress-strain curves constructed using the strain data recorded by the extensometer. In order to obtain consistent test results, at least three repeat tests were conducted for each material at each speed at RT and 77 K. For 20 K tests, due to the cost of liquid helium and long cooling time, at least two repeat tests were performed to obtain consistent stress-strain curves.

2.2 The set-up of thermal expansion coefficient measurements

The CTE coupons were prepared according to the ASTM D696-16 [19]. This standard does not make specific reference to conducting tests at cryogenic temperature, but has been adapted to operate in this temperature range. Cylindrical test coupons with a nominal diameter of 13 mm and nominal height of 55 mm were produced. A detailed view of the CTE measurement rig and setup is shown in Figure 4 [2]. The rig consists of a double-walled vacuum insulated silver-coated glass Dewar sitting vertically inside a foam bucket to create a 77 K liquid bath around the outside wall of the glass Dewar. This assembly is located inside a large outer container lined with Multilayer insulation to further minimise heat ingress.

The dilatometer consists of a Quartz glass tube and hollow glass cylinder. Quartz has negligible thermal expansion in the range of 300 K to 4 K. The glass tube and cylinder are the same length and the cylinder is able to slide freely inside the tube. The sample was inserted into a copper cylinder with outer diameter matching the bore of the glass tube and its centre bored with the hole to suit the sample cylinder. A calibrated silicon diode thermometer was glued into a groove in the copper sample holder to measure its local temperature. The sample holder with sample was placed at the bottom glass tube and the glass cylinder placed on top. The glass cylinder follows the movement of the sample as it expands and contracts. A linear variable differential transformer (LVDT) device was positioned at the top of the assembly. The "pin" or moving core of the LVDT was fixed directly to the glass cylinder, and the "outer winding" of the LVDT clamped in a fixed position. As the pin moves up and down inside the outer winding, a change in voltage is recorded that corresponds with the changing height of the sample. The LDVT has a working range of ± 4 mm, where the corresponding change in voltage recorded is linear.

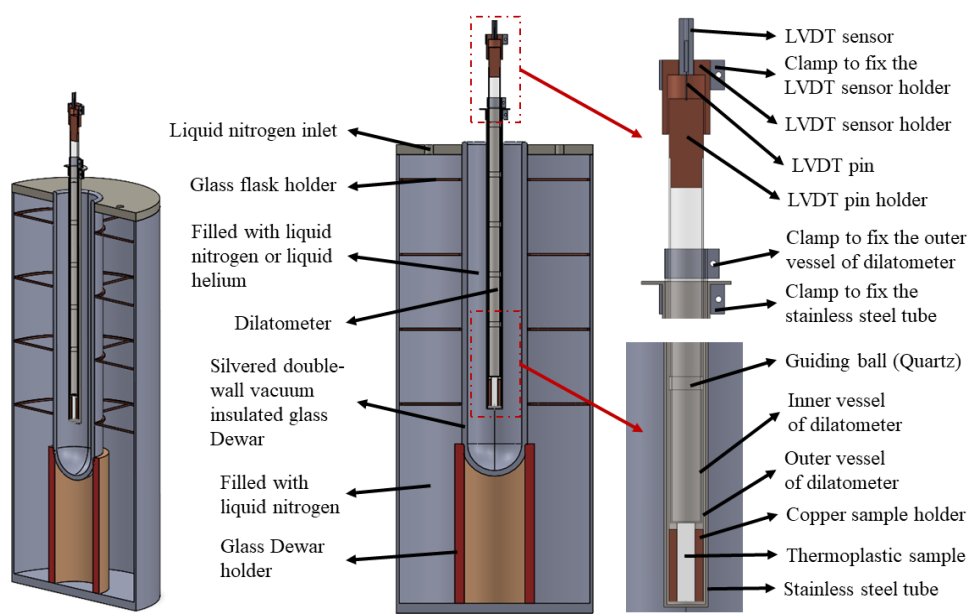

Figure 4. The illustration of bespoke designed CTE measurement rig [2].

The CTEs of the same three polymers were measured from 20 K to RT (295 K). Once a sample was installed, the LVDT was adjusted to ensure it was measuring within its linear range and started close enough to 0 mm. Liquid nitrogen was filled into the foam bucket and liquid helium was transferred into the glass Dewar until the sample reached the target temperature. The liquid helium and liquid nitrogen gradually evaporated and the sample temperature slowly increased back to RT after 20 hours. Due to the slow warming up, quasi thermal equilibrium could be achieved between 20 K and RT. The LVDT recorded the amount of expansion of sample at each temperature from 20 K to RT and the LVDT voltage was converted to millimetres. The linear CTEs, α, between temperatures $T_1$ and $T_2$ were calculated by using the Equation 1,

$$\alpha = \frac{L_2 - L_1}{L_0(T_2 - T_1)}, \qquad (1)$$

where $L_0$ is the original length of sample at RT. $L_1$ and $L_2$ are the length of the sample respectively at temperatures $T_1$ and $T_2$. For each of the three materials; two repeat measurements were conducted.

2.3 Scanning electron microscope

To understand the characteristics and fracture mechanism of thermoplastic polymers under tension, the fracture surfaces of PTFE, PEEK and UHMWPE coupons after all tensile tests were examined with a JEOL JSM-5910 Scanning Electron Microscope (SEM) at a scale of 25 μm. Before observing the fracture surface, the fracture surface was cleaned by using ethanol and the fracture surface was sputter coated with a gold (Au) powder to add the conductive surface required for SEM examination. The thickness of Au-coating was ~20 nm.

3. Results

3.1 Temperature and strain rate effects on tensile properties

The detailed mechanical properties of PTFE, PEEK and UHMWPE at each strain rate are summarised in Table 1 for RT, Table 2 for 77 K, and Table 3 for 20 K. The stress-strain curves of corresponding repeat tests are illustrated in Figures 1s, 2s and 3s of supplementary information. The individual stress-strain curve most representative of the batch is selected (using the standard deviation to the mean value for the batch) to construct further plots in Figures 5, 6 and 7 for direct comparisons of the tensile performance with respects to the changes in strain rate, Young's modulus, and UTS for each test temperature.

Table 1. Strain rate effects at RT

|  | Mean Young's modulus, $E_t$ (GPa) [Std. Dev] | Mean UTS, $\sigma_m$ (MPa) [Std. Dev] | Mean nominal strain at UTS, $\varepsilon_{tm}$ (%) [Std. Dev] | Mean nominal strain at break, $\varepsilon_{tb}$ (%) [Std. Dev] |
|---|---|---|---|---|
| PTFE |  |  |  |  |
| 2 mm/min | 0.82 [0.02] | 18.15 [0.49] | 301.59 [17.23] | 302.96 [17.31] |
| 10 mm/min | 1.04 [0.03] | 19.23 [0.41] | 249.97 [12.44] | 252.56 [11.67] |
| 50 mm/min | 1.04 [0.02] | 19.94 [0.73] | 202.93 [10.84] | 207.24 [9.91] |
| 250 mm/min | 1.14 [0.02] | 19.23 [0.15] | 115.88 [5.94] | 117.82 [6.86] |
| PEEK |  |  |  |  |
| 2 mm/min | 3.71 [0.02] | 99.55 [2.72] | 14.46 [0.13] | 29.00 [1.01] |
| 10 mm/min | 3.69 [0.04] | 98.45 [0.80] | 10.59 [0.60] | 25.83 [1.05] |
| 50 mm/min | 3.71 [0.04] | 102.96 [1.01] | 9.94 [0.12] | 20.91 [1.21] |
| 250 mm/min | 3.75 [0.02] | 107.28 [2.35] | 9.89 [0.30] | 18.63 [1.12] |
| UHMWPE |  |  |  |  |
| 2 mm/min | 0.87 [0.03] | 31.99 [0.45] | 318.76 [15.59] | 329.13 [13.00] |
| 10 mm/min | 0.94 [0.04] | 32.61 [0.82] | 305.54 [10.88] | 316.64 [6.98] |
| 50 mm/min | 1.00 [0.02] | 34.67 [0.27] | 288.63 [13.59] | 295.43 [13.83] |
| 250 mm/min | 1.02 [0.03] | 34.77 [0.26] | 244.70 [21.51] | 255.50 [17.76] |

Table 2. Strain rate effects at 77 K

|  | Mean Young's modulus, $E_t$ (GPa) [Std. Dev] | Mean UTS, $\sigma_m$ (MPa) [Std. Dev] | Mean strain at UTS and break, $\varepsilon_m = \varepsilon_b$ (%) [Std. Dev] |
|---|---|---|---|
| PTFE |  |  |  |
| 2 mm/min | 5.71 [0.19] | 86.22 [0.23] | 4.44 [0.08] |
| 10 mm/min | 5.88 [0.06] | 89.87 [1.31] | 4.42 [0.17] |
| 50 mm/min | 6.07 [0.22] | 90.41 [1.69] | 4.04 [0.22] |
| 250 mm/min | 6.13 [0.13] | 89.11 [1.72] | 3.38 [0.24] |
| PEEK |  |  |  |
| 2 mm/min | 5.32 [0.14] | 198.14 [1.33] | 8.09 [0.41] |
| 10 mm/min | 5.43 [0.12] | 212.34 [2.87] | 6.13 [0.16] |
| 50 mm/min | 5.47 [0.12] | 211.26 [6.87] | 5.96 [0.30] |
| 250 mm/min | 5.47 [0.04] | 214.72 [5.12] | 5.08 [0.19] |
| UHMWPE |  |  |  |
| 2 mm/min | 6.63 [0.04] | 129.77 [0.70] | 2.60 [0.13] |
| 10 mm/min | 7.81 [0.45] | 136.90 [0.67] | 2.59 [0.06] |
| 50 mm/min | 7.45 [0.14] | 141.03 [1.45] | 2.58 [0.07] |
| 250 mm/min | 7.67 [0.20] | 139.02 [0.10] | 2.28 [0.11] |

Table 3. Strain rate effects at 20 K

|  | Mean Young's modulus, $E_t$ (GPa) [Std. Dev] | Mean UTS, $\sigma_m$ (MPa) [Std. Dev] | Mean strain at UTS and break, $\varepsilon_m = \varepsilon_b$ (%) [Std. Dev] |
|---|---|---|---|
| PTFE |  |  |  |
| 2 mm/min | 6.24 [0.04] | 82.32 [0.59] | 2.51 [0.04] |
| 10 mm/min | 6.77 [0.12] | 81.23 [1.71] | 2.36 [0.04] |
| 50 mm/min | 7.27 [0.23] | 93.21 [1.86] | 2.25 [0.23] |
| 250 mm/min | 6.47 [0.11] | 92.78 [1.26] | 2.26 [0.07] |
| PEEK |  |  |  |
| 2 mm/min | 5.61 [0.19] | 114.57 [2.14] | 2.70 [0.06] |
| 10 mm/min | 5.63 [0.10] | 123.31 [3.31] | 2.57 [0.08] |
| 50 mm/min | 6.03 [0.04] | 135.43 [1.05] | 2.36 [0.02] |
| 250 mm/min | 5.81 [0.10] | 133.26 [1.09] | 2.45 [0.03] |
| UHMWPE |  |  |  |
| 2 mm/min | 7.68 [0.29] | 99.48 [2.88] | 1.44 [0.01] |
| 10 mm/min | 8.51 [0.20] | 105.06 [0.25] | 1.39 [0.07] |
| 50 mm/min | 8.88 [0.10] | 105.07 [2.68] | 1.29 [0.03] |
| 250 mm/min | 8.98 [0.21] | 98.69 [0.19] | 1.18 [0.02] |

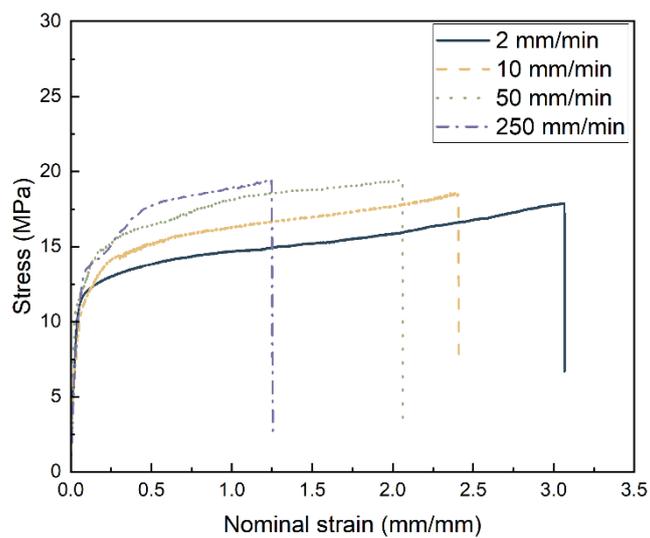

(a)

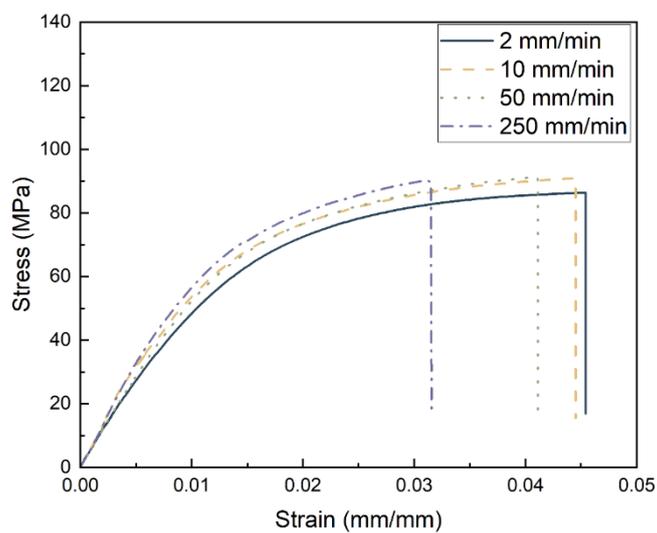

(b)

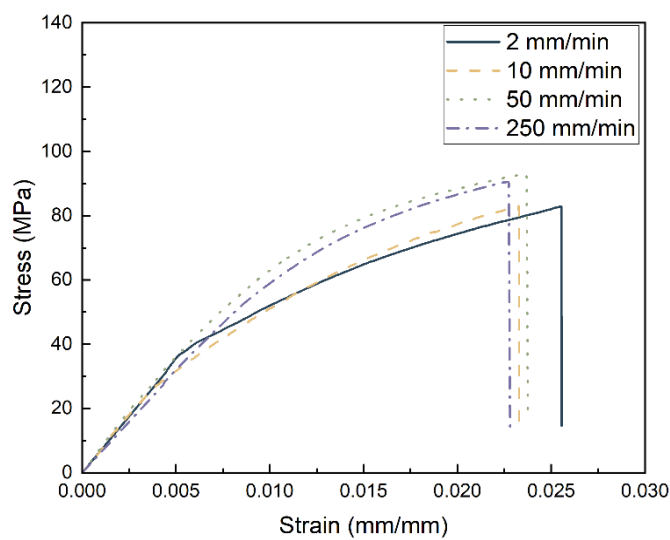

(c)

Figure 5. Comparison of strain effects on PTFE at: (a) RT; (b) 77 K; (c) 20 K.

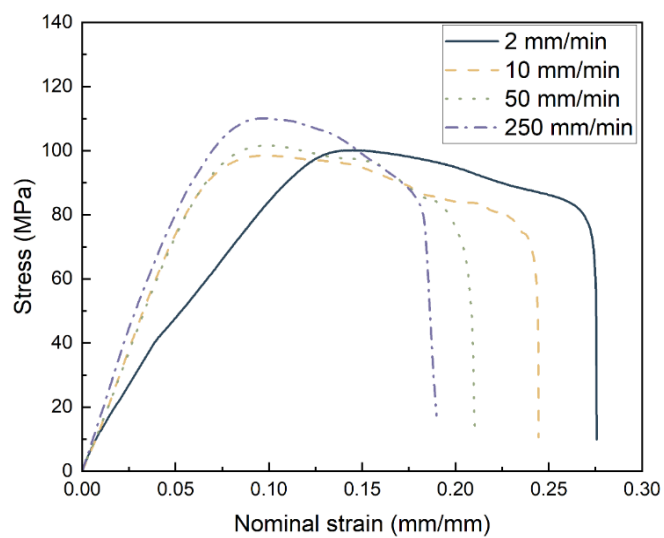

(a)

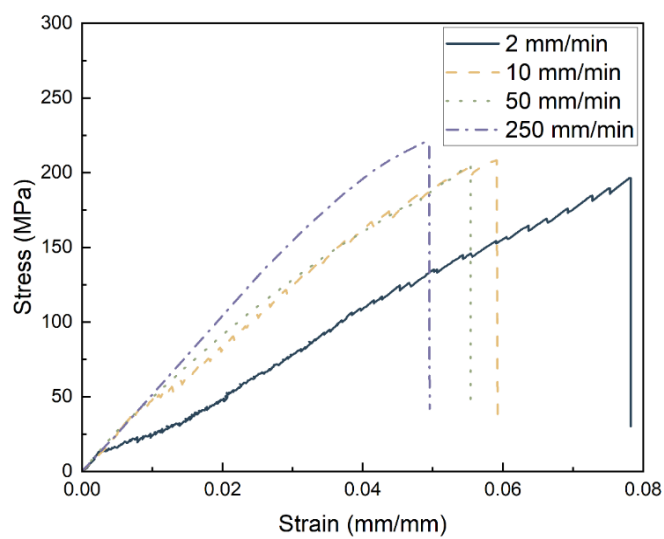

(b)

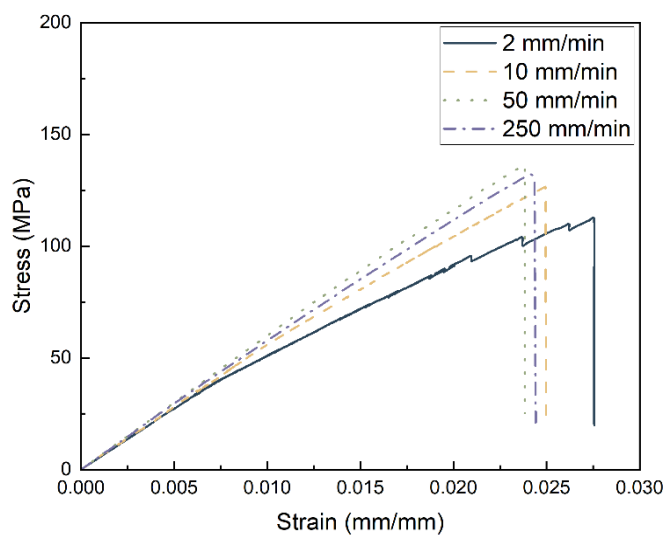

(c)

(c)

Figure 6. Comparison of strain effects on PEEK at: (a) RT; (b) 77 K; (c) 20 K.

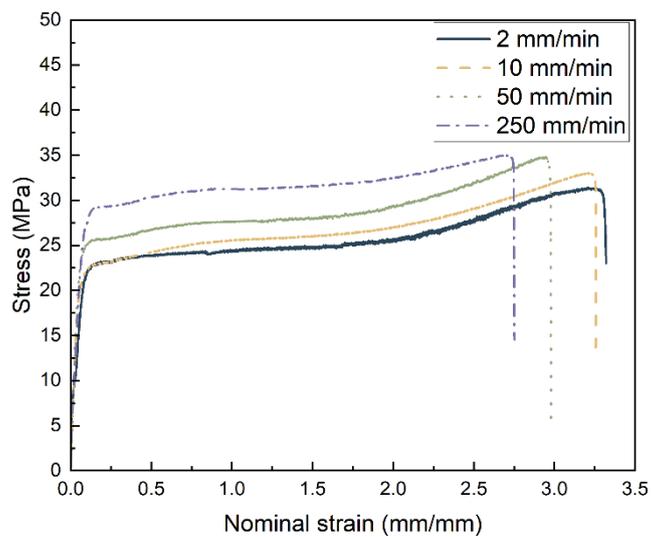

(a)

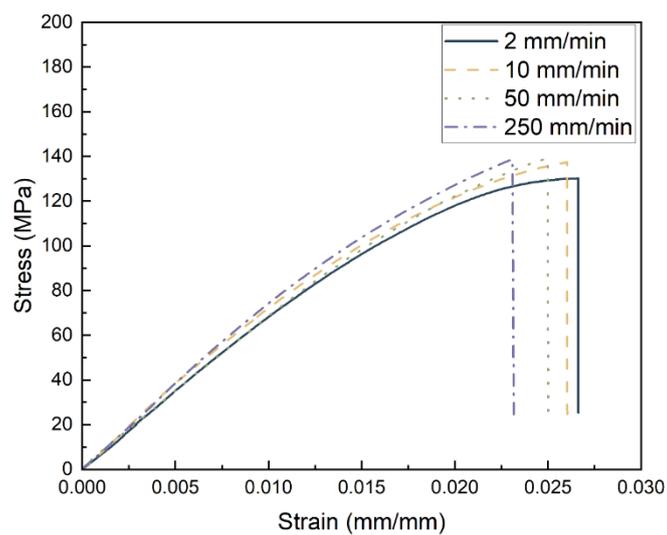

(b)

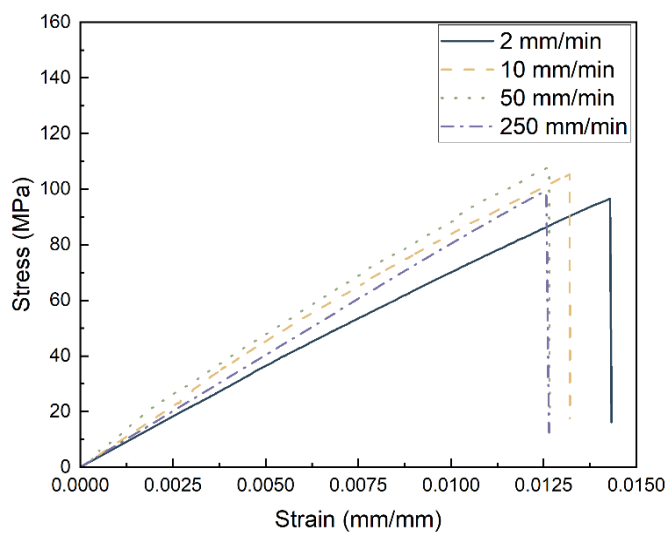

(c)

Figure 7. Comparison of strain effects on UHMWPE at: (a) RT; (b) 77 K; (c) 20 K.

The three thermoplastic polymers have their own unique advantages for particular applications. At ambient environment, PEEK has the highest Young's modulus and UTS at all four strain rates when compared to the other thermoplastics. The Young's modulus was over 3 times higher than PTFE and UHMWPE. In contrast, the UTS of PTFE and UHMWPE were ~20 % and ~34 % smaller than the maximum of UTS of PEEK and the nominal strains were at least 12 and 22 times larger than those of PEEK at all four strain rates. Similarly, the nominal strains at break of the same two polymers were at least 6 and 11 times higher than those of PEEK.

At 77 K, PEEK still had the highest UTS, however, it recorded the lowest Young's modulus among them, while UHMWPE had the highest modulus. In contrast to the behaviour found in RT tests, PEEK had the highest strain at UTS and at break when compared to the other two polymers at 77 K. UHWMPE had the smallest strain at UTS and at break at each of the four strain rates.

At 20 K, in general, the relationships and trends reported were similar to those at 77 K. PEEK still had the highest UTS at 20 K, while PTFE had the lowest. However, at this temperature, PEEK's overall UTS performance margin over UHMWPE and PTFE significantly reduced by 19 % and 97 % respectively. At 20 K, UHMWPE had the highest stiffness while PEEK had the lowest. It was also found that PTFE and PEEK had similar strains at both UTS and at break at this temperature. The mean strain at UTS reported for PTFE and PEEK was ~2.4%, while UHMWPE had a much lower strain of 1.3%.

The effect of cryogenic temperature significantly changed the mechanical behaviours of the three types of thermoplastics. It is explicit that the strains measured at UTS significantly reduced when the temperature was reduced from 295 K to 20 K. For PTFE measured at a speed 2 mm/min, the strain at UTS reduced by 99 % at 77 K, when compared to the same strain reported at RT. The strain at UTS further reduced by 44 % when the temperature reduced to 20 K. For PTFE at 250 mm/min, the strain at UTS was reduced by 97 % at 77 K when compared to its nominal strain at UTS at RT. This value further reduced by 33%, when the temperature reduced to 20 K. These patterns were also found for the other two polymers.

The Young's modulus of all three thermoplastic polymers increased significantly at 77 K. The maximum increases reported were 47 %, 596 % and 731 % for PEEK, PTFE and UHMWPE respectively when compared to the equivalent RT results. Large increases in their UTS were also reported when temperature was reduced to 77 K, especially for PTFE (maximum 375 %) and UHMWPE (maximum 320 %). However, when temperature was reduced to 20 K, only marginal increases in the Young's modulus were discovered and the UTS for PEEK and UHMWPE both decreased by 42 % and 29 % respectively. The only divergence from this trend was a very modest increase of UTS of ~4 % measured for PTFE at the higher strain rates of 50 mm/min and 250 mm/min at 20 K.

To summarise, when the strain rate was increased, both Young's modulus and UTS for all material also increased, but the strain tolerated before the material failed tended to decrease. In comparison, the Young's modulus and UTS did not change when strain rate was > 10 mm/min at 77 K. However, the maximum strain recorded at both UTS and at break (failure) did decrease with increasing strain rate. When the temperature was reduced to 20 K, these patterns tended to continue with only some subtle changes in performance.

PTFE showed an increase in UTS and a reduction in its strain at both UTS and at break for increasing strain rate. However, the Young's modulus of PTFE began to decrease for strain rates over 50 mm/min. The same behaviour was also found for PEEK. For UHMWPE, the Young's modulus and the strains at UTS and at break followed the same trend as those at RT, and 77 K with respects to increasing strain rates. However, its UTS was not influenced by increasing strain rate.

3.2 Thermal expansion coefficients from 20 K to room temperature

Figure 8 illustrates the percentage contraction (change in length over original length) of PTFE, PEEK and UHMWPE between 20 K and 295 K. The contraction of polymers was found to be non-linear and therefore, the linear CTEs were calculated within different temperature ranges and listed in Table 4. The contraction for all the materials tended to plateau below ~ 50 K. When the temperature was at 20 K, PTFE, PEEK and UHMWPE contracted by ~1.7%, 0.9% and 1.9% respectively, with PEEK contracting much less than the other polymers across the full temperature range.

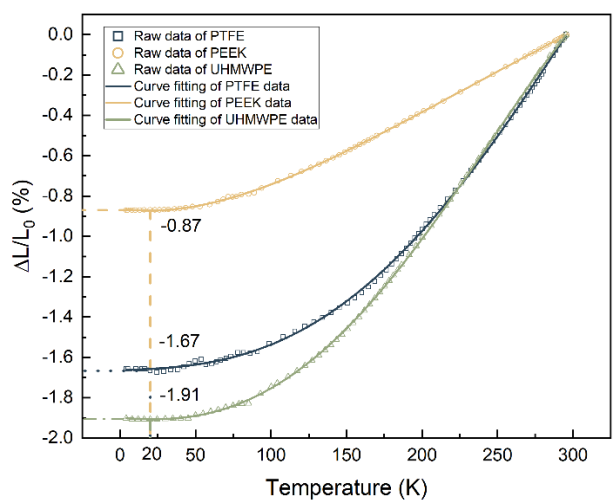

Figure 8. The relative thermal contraction of PTFE, PEEK and UHMWPE.

Table 4. Summary of linear coefficients of thermal expansion

|  | Mean CTE, $\alpha$ ($10^{-6}$/K) [Std. Dev] |
|---|---|
| PTFE |  |
| 20 – 50 K | 8.28 [1.85] |
| 50 – 123 K | 25.28 [1.00] |
| 123 – 200 K | 62.31 [0.26] |
| 200 – 295 K | 103.06 [0.04] |
| PEEK |  |
| 20 – 50 K | 6.15 [1.19] |
| 50 – 123 K | 23.47 [1.04] |
| 123 – 200 K | 37.72 [0.25] |
| 200 – 295 K | 40.76 [0.59] |
| UHMWPE | |
| 20 – 50 K | 7.01 [1.91] |
| 50 – 123 K | 36.02 [2.41] |
| 123 – 200 K | 81.33 [1.12] |
| 200 – 295 K | 106.81 [0.18] |

3.3 Microscopic analysis

The observations on the fracture surfaces of PEEK, PTFE and UHMWPE by using SEM are presented in this section.

Figures 9 to 11 show the tensile fracture surfaces of PEEK at RT, 77 K and 20 K from tests conducted at four different crosshead speeds. By comparing the four sets of images, the evolution of the fracture surface roughness was found to increase with decreasing temperature, while an increase in the number and density of "roughness lines" flowing in a common direction indicate greater deflections occurring on approach to failure [20]. Observing the fracture surfaces from RT tests as shown in Figure 9, the overall surfaces remained smooth. No facet features could be found; however, remarkable plastic deformations (marked by white circles) were seen towards the edges of the samples. The overall textured micro-flows are generally in parallel. From the fractured surfaces obtained after cryogenic mechanical tests in Figures 10 and 11, the scarps and textured micro-flow generally propagate radially from defects, which are marked by yellow circles to identify the defects and by red lines to indicate the direction of propagation. Plastic deformation is hardly ever found from these surfaces. Comparing the fractography obtained from 77 K and 20 K tests, the 20 K fractures show more facet surfaces, and have greater crack deflections. Additionally, the instantaneous fracture area is explicitly increased, and the number of mirror-like and opaque ("misty") areas decreases when the temperature is reduced from 77 K to 20 K [21].

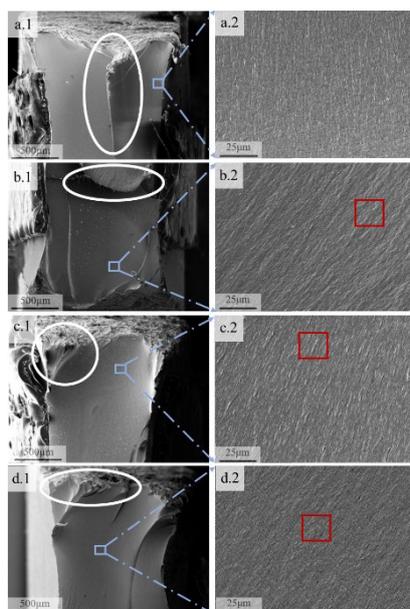

Figure 9. SEM images of the tensile fracture morphologies of PEEK at RT: (a.1)(a.2) 2 mm/min; (b.1)(b.2) 10 mm/min; (c.1)(c.2) 50 mm/min; (d.1)(d.2) 250 mm/min.

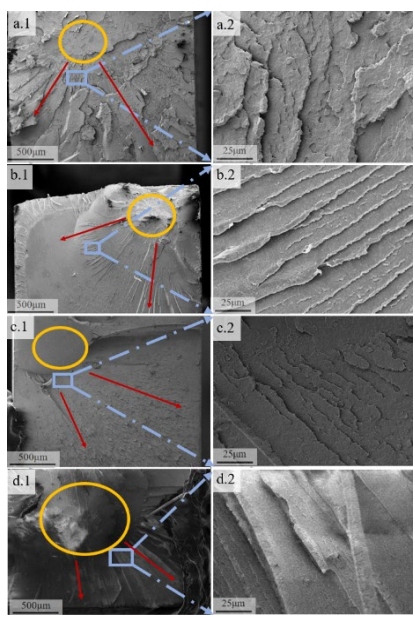

Figure 10. SEM images of the tensile fracture morphologies of PEEK at 77 K: (a.1)(a.2) 2 mm/min; (b.1)(b.2) 10 mm/min; (c.1)(c.2) 50 mm/min; (d.1)(d.2) 250 mm/min.

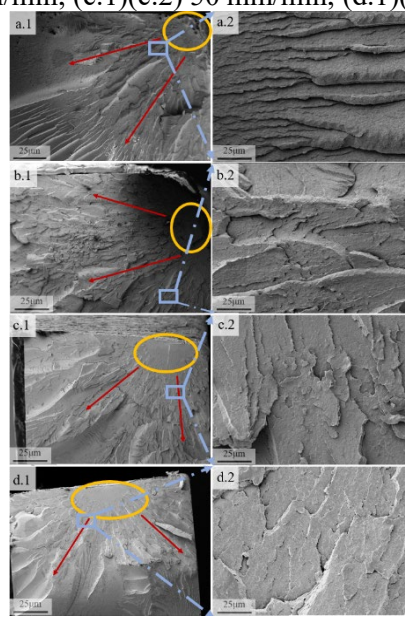

Figure 11. SEM images of the tensile fracture morphologies of PEEK at 20 K: (a.1)(a.2) 2 mm/min; (b.1)(b.2) 10 mm/min; (c.1)(c.2) 50 mm/min; (d.1)(d.2) 250 mm/min.

Because the effect of strain rates on the form and density of emerging features remained inconclusive, the fracture surfaces of PEEK samples were taken at a higher magnification in Figures 9 to 11. For the samples tested at RT, the fracture surface is relatively smooth and uniform at 2 mm/min. When the strain rate is increased to 10 mm/min, the surface develops little scarps (marked by red rectangle in Figure 9 b.2). At 50 mm/min, the fracture surface evolves and has more cusps (marked by red rectangle in Figure 9 c.2). Finally, when the strain rate is increased to 250 mm/min, the cleavage steps become more explicit (marked by red rectangle in Figure 9 d.2) [22]. One emerging trend is for the fracture surfaces obtained from samples tested at higher strain rates at RT to tend to have an increasing surface roughness. For samples tested at 77 K and 20 K, the fracture surface remains rather similar up until the strain rate exceeds 10 mm/min where there is some demarcation. The surfaces are highly multi-faceted at 2 mm/min, and become increasingly smoother as the strain rate increased to 250 mm/min.

Figure 12 shows the fracture surfaces of PTFE samples tested at RT, 77 K and 20 K at the four different speeds. Similar to PEEK, the fracture surfaces of PTFE from RT tests are significantly different to those obtained after the cryogenic tests. The microscope images taken after RT test are shown in Figures 12a, 12d, 12g and 12j. Textured linear lines showing the direction of micro-flows (marked by blue arrows) are visible, and the overall surfaces are smooth, which are similar to the observations found for PEEK. When comparing the RT fractures, with those obtained from the cryogenic tests (shown in the second and third columns of Figure 12), the latter are visually much rougher, and a number of circular lamellas crystal nucleus were found (marked by yellow circles). The cracks (marked by red arrows in Figure 12) in every lamella propagate radially from its crystal nucleus. When comparing the surfaces between 77 K and 20 K, the distances between lamellas crystal nuclei (distance between yellow circles) of 77 K samples, are shorter than those found in 20 K samples.

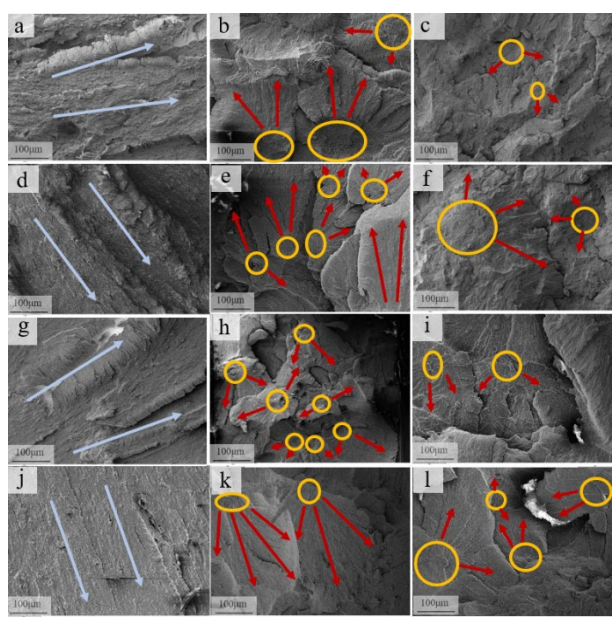

Figure 12. SEM images of the tensile fracture morphologies of PTFE at :(a) RT 2 mm/min; (b) 77 K 2 mm/min; (c) 20 K 2 mm/min; (d) RT 10 mm/min; (e) 77 K 10 mm/min; (f) 20 K 10 mm/min; (g) RT 50 mm/min; (h) 77 K 50 mm/min; (i) 20 K 50mm/min; (j) RT 250 mm/min; (k) 77 K 250 mm/min; (l) 20 K 250 mm/min.

Figure 13 shows the fracture surfaces of PTFE at higher magnification, which were used to investigate the strain effects on the fracture surfaces. It was found in the two sets of cryogenic fracture surfaces photos (Figures 13b, 13e, 13h, 13k and 13c, 13f, 13i, 13l) that the microstructures do not significantly change with increased strain rate. However, for RT fractures, smaller dimples in lower densities emerge as the strain rate was increased.

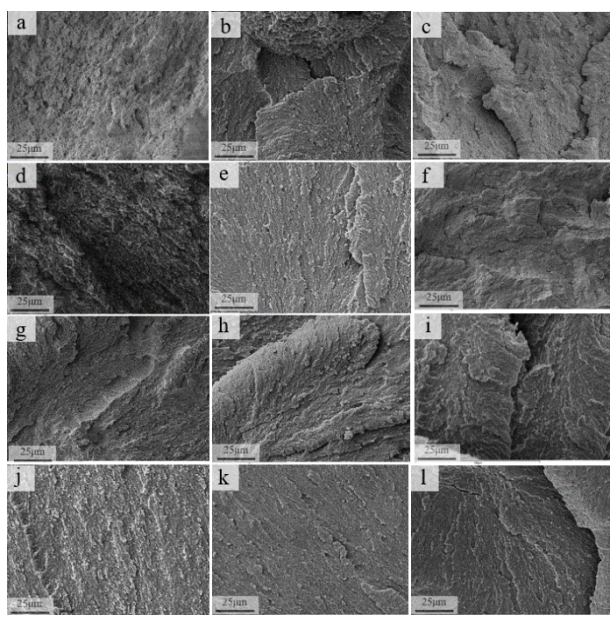

Figure 13. SEM images of the tensile fracture morphologies of PTFE at :(a) RT 2 mm/min; (b)77 K 2 mm/min; (c) 20 K 2 mm/min; (d) RT 10 mm/min; (e) 77 K 10 mm/min; (f) 20 K 10 mm/min; (g) RT 50 mm/min; (h) 77 K 50mm/min; (i) 20 K 50mm/min; (j) RT 250 mm/min; (k) 77 K 250mm/min; (l) 20 K 250mm/min.

Figure 14 shows that temperature has a significant evolution effect on the fracture surface of UHMWPE when comparing the three columns of images. For samples tested at RT, the fracture surfaces shown in Figures 14 (a, d, g and j) illustrates that plastic deformation (marked by white circles) is apparent, and the cracks (marked by red arrows) are linear. However, in the next two columns of Figures 14 (b, e, h, k) and 14 (c, f, i, l), the fracture surfaces obtained after cryogenic testing (both 77 K and 20 K) show defined facets. A "crocodile-skin-like" surface populated with many crack deflections texture was characteristic of the cryogenically failed samples. Figures 14b and 14e show good examples at 77 K, and indicate how the cracks have initiated from the "flat mirror-like" areas in yellow circles, and propagated radially in the direction of the red arrows. In contrast, Figures 14 (c, f, i, l) obtained from the 20 K tests, shows evidence of crack deflection with smaller angles, and the single facet areas are larger than those at 77 K. Moreover, the size and population of "flat mirrored" and "misty" areas found were greatly decreased denoting even more instantaneous fracturing had occurred at this temperature [21].

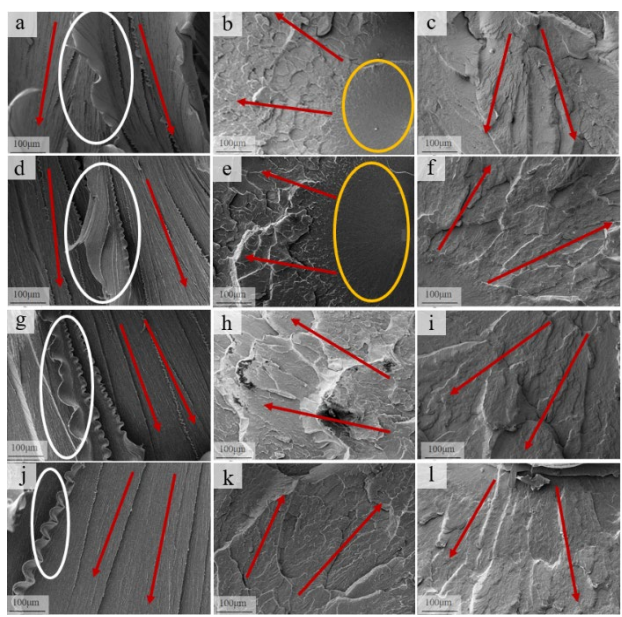

Figure 14. SEM images of the tensile fracture morphologies of UHMWPE at: (a) RT 2 mm/min; (b) 77 K 2 mm/min; (c) 20 K 2 mm/min; (d) RT 10 mm/min; (e) 77 K 10 mm/min; (f) 20 K 10 mm/min; (g) RT 50 mm/min; (h) 77 K 50mm/min; (i) 20 K 50 mm/min; (j) RT 250 mm/min; (k) 77 K 250mm/min; (l) 20 K 250 mm/min.

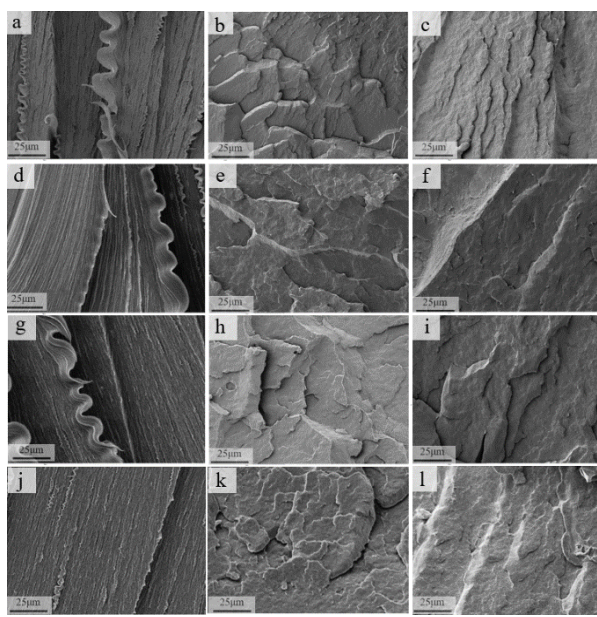

Figure 15 SEM images of the tensile fracture morphologies of UHMWPE at :(a) RT 2 mm/min; (b)77 K 2 mm/min; (c) 20 K 2 mm/min; (d) RT 10 mm/min; (e) 77 K 10 mm/min; (f) 20 K 10 mm/min; (g) RT 50 mm/min; (h) 77 K 50mm/min; (i) 20 K 50mm/min; (j) RT 250 mm/min; (k) 77 K 250mm/min; (l) 20 K 250mm/min.

Figure 15 uses higher magnification photos to show the strain rate effects on the fracture surfaces of UHMWPE by comparing images in the horizontal rows. Overall, the influence of strain rate was not explicit at cryogenic temperatures. For RT tested samples, the increased strain rate slightly reduced the plastic deformation, which is reflected by the reduced heights of the plastic deformation strips.

4. **Discussion**

By combining the results and observations from Figures 5 to 7 and Figures 9 to 15; lower temperature and higher strain rates reduce the ductility and increase the brittleness in all three materials. The tensile strengths of the materials are significantly enhanced when the temperature is reduced from RT to cryogenic temperatures. One of the main reasons is connected to their glass transition temperature. This threshold temperature defines the change in a polymer's ability to behave "rubbery and ductile" to "glassy and rigid" [23]. The low temperature constrains the mobility of the polymer molecules and chains, and stores up fracture energy by resisting plastic flow. Thus, low temperature will induce an increase in the materials strength. However, at some instance, cleavage-like cracks will start to emerge once the stored energy reaches a limit [24]. The changes in facet fracture formation and texture can be clearly observed under SEM. Furthermore, the glass transition temperatures of polymers are different. The glass transition temperature for PEEK is 416 K, for PTFE is170 K, and for UHMWPE is123 K [25–27]. Therefore, only PEEK has a glass transition at RT, which is the reason why PEEK has higher strength and lower ductility than PTFE and UHMWPE at RT. Conversely, when PTFE and UHMWPE are tested at 77 K and 20 K, (i.e. temperature lower than their glass transition temperatures), they both show significantly increased strength and decreased ductility performance compared to PEEK.

It is understood that fracture surface morphologies demark the differences in strength that emerge between different thermoplastics at cryogenic temperatures. For example, both PEEK and UHMWPE form large size hackle patterns, but PTFE forms spherulites. The presence of the hackle pattern implies the polymer has better continuity and ability to distribute stress during fracture. Therefore, PEEK and UHMWPE have higher strength than PTFE at cryogenic temperatures. These findings are consistent with the findings in [27–29]. Furthermore, the cracks on PEEK's fracture surface have less deflections than those of UHMWPE. This indicates the stress concentration forming in PEEK is less than in UHMWPE. Thus, PEEK has higher strength. These phenomena are consistent with our mechanical tests results shown in Tables 2 and 3 and Figures 5 to 7.

When the test temperature is decreased from 77 K to 20 K, the Young's modulus of these thermoplastic marginally increase. Their reported UTS values decrease and so to the maximum strain recorded at UTS. It reflects that molecule movement is becoming significantly restricted, and ductility is starting to plateau, leading to the onset of embrittlement on approach to 20 K. This pattern is consistent with the observations from Figures 10, 11, and 14. The discovery of mirror-like areas reduce, and instantaneous fracture areas increase in both PEEK and UHMWPE when the temperature is reduced from 77 K to 20 K. This indicates the fracture speeds also increases at 20 K, driven mainly by the increased susceptibility to embrittlement [22, 31]. Furthermore, the reduced UTS is consistent with the increased facet fracture surface. Comparing 77 K to 20 K results; about 80% of PEEK's fracture surface is changed from "mirror-like" or opaque (misty) areas to instantaneous fracture areas, and correspondingly its UTS decreases by nearly 50%. Over the same change in temperature, only 7% of UHMWPE's fracture surface changed into instantaneous fracture area, and its UTS is decreased by over 20%. Moreover, for PEEK and UHMWPE, the crack tip deflection angle is smaller at 20 K compared to 77 K. It is speculated they are also influenced by the increased fracture speed.

The sudden and unexpected stress drops found in Figure 6 for the PEEK tested at 2 mm/min at cryogenic temperatures could be caused by an instant fracture. PEEK has long, randomly distributed and strong molecular chain. When it is under tension, uneven stress on molecular chains causes early failure of some chains. When a large area facet emerges, its existence becomes symbolic in the stress-strain curve by an increase in the number of sudden stress drops occurring during the trace.

For PTFE, according to Ross et al. [28], spherical crystallization can cause voids between crystals, which leads to stress concentration. High cooling rate generates stacked crystals. This is because rapid cooling increases the concentration of connected molecules, and long molecules are more easily folded into two layers when a significant amount of layers are formed within a short time period. These inter-crystalline links may not be as tight as those formed by slow crystallisation [28]. This suggest the distance between lamellas crystal nucleus in PTFE at 77 K are shorter than those found at 20 K since 77 K tests since it can be considered the rate of cooling at 77 K was more rapid due to direct liquid. It is considered that adding nanoparticles into these polymers could be a promising solution to reduce the stress concentration caused by rapid cooling since they can reduce the CTE [31]. In addition, to improve the mechanical behaviours of these polymers, the optimisation of polymer manufacturing process is critical. If the crystallization position and direction can be shaped and the defects are reduced, the strength of them can be greatly enhanced. Zone annealing could be a solution since it can tightly control the thickness of semi crystalline polymers by directionally solidifying the material under optimal conditions [32].

The thermal expansion of polymers is determined by various modes of thermal vibration propagating along and between the polymer chains, and influenced by the glass transition process [33]. At cryogenic temperatures, it is known that thermal contraction is dominated by the non-harmonic oscillation due to the asymmetry of internal binding potentials; when the temperature approaches 0 K, the binding potentials between vibrating units are symmetric, since the asymmetric higher-energy levels of these oscillators are restrained [33]. Therefore, the CTEs of them all tend to reduce to zero with reduced temperature. For PEEK, it remains in its glassy state over the measured temperature range in this investigation (295 K – 20 K) since its glass transition temperature (416 K) is higher than room temperature. This means its thermal expansion behaviours are only controlled by the inter/intrachain vibration modes, and not affected by glass transition at the temperature below 295 K.

In Figure 8, when temperature is above 50 K, the CTE of PEEK clearly increases with rising temperature, which may be caused by the increasing dominance of three-dimensional interchain vibrations in this temperature range. When the temperature is increased, the interchain vibration tends to become saturated, and thermal expansion is no longer sensitive to the change of temperature [33], and the thermal expansion curve becomes linear. For both PTFE and UHMWPE, their glass transition temperatures are much lower (170 K & 123 K respectively). This leads to the increase in free volume available for vibration, resulting in a significant rise in their CTEs around the transition temperature. The lower glass transition temperature of UHMWPE compared to PTFE could potentially contribute to its own ability to contract more at cryogenic temperature; although the effect of glass transition upon thermal expansion is rather qualitative than quantitative [33]. Therefore, to adjust the thermal expansion behaviours of thermoplastic polymers; modifying their internal free volume and changing their glass transition temperature are considered effective approaches, including adding reinforcement fillers such as nanoparticles or fibres [10,34,35]. In many applications, a particular rate of contraction with respect to temperature is normally required to fine tune the operation of the components such as valves. Polymers with relatively low CTEs would work effectively to avoid development of stress concentrations caused by thermal mismatches between metals [36,37]. While in systems operating dynamically over the cryogenic temperature range, those polymers with higher thermal contraction or expansion can be installed initially under interference fit [38] to ensure after thermo-mechanical loading down to low temperature they continue to seal effectively [39].

## 5. Conclusion

This study measured the thermal expansion behaviour and tensile properties of three types of popular thermoplastic polymers at RT, 77 K and 20 K at different strain rates. The key findings are:

- the UTSs of the three types materials were all increased at 77 K and were mostly deceased when the temperature was further reduced to 20 K across the four strain rates;
- Young's moduli increased and strains at UTS decreased with the reduced temperature from RT to 20 K;
- At RT, a higher strain rate resulted in reduced plastic deformation at the fracture surface, while at cryogenic temperatures, the surface morphology exhibited an insensitivity to increasing strain rate;
- For PEEK and UHMWPE, a reduction in temperature from 77 K to 20 K resulted in a larger instantaneous fracture area, and a multi-faceted fracture surface containing smaller mirror-like and opaque (misty) sub-regions. In the case of PTFE, the dimple sizes become smaller at 20 K when compared to those observed at 77 K;

- Of the three polymers investigated, PEEK contracts least at 20 K. PTFE and UHMWPE have much lower glass transition temperatures, which partly contributes to them contracting much more at cryogenic temperatures.

**Acknowledgements**

This project was supported by the UKRI STFC Grant (Grant No. ST/T001844/1). In addition, technical support from Mr Rob Loades and Mr Owain Atkins and facilities from the Institute of Cryogenics, University of Southampton, Southampton, UK are gratefully acknowledged. We thank Mr Hafizuddin Bin Hassim from University of Southampton for assisting the experimental work.